\documentclass[10pt,english,preprintnumbers,superscriptaddress,floatfix,
showpacs,prd,aps,nofootinbib]{revtex4}
\usepackage{amssymb}

\usepackage[T1]{fontenc}
\usepackage[latin9]{inputenc}
\usepackage{amsmath}
\usepackage{graphicx}
\usepackage{hyperref}
\usepackage{babel}

\newcommand{\ave}[1]{\left\langle #1 \right\rangle}

\begin{document}

\title{Acceptance corrections to net baryon and net charge cumulants}
\author{Adam Bzdak}
\email[E-Mail:]{ABzdak@bnl.gov}
\affiliation{RIKEN BNL Research Center, Brookhaven National Laboratory, Upton, NY 11973,
USA}
\author{Volker Koch}
\email[E-Mail:]{VKoch@lbl.gov}
\affiliation{Lawrence Berkeley National Laboratory, Berkeley, CA 94720, USA}
\pacs{25.75.-q, 24.85.+p, 24.60.-k}
\date{\today }

\begin{abstract}
We show that the effect of finite acceptance drastically influences the
net-baryon and net-charge cumulants, which are believed to be sensitive
probes of the QCD phase diagram. We derive the general formulae that relate
the true cumulants $K_{n}$ which reflect the full dynamics of the system
with the actually measured cumulants $c_{n}$ for a given acceptance,
modeled by a binomial probability parameter $p$. We find that this relation
involves additional moments which cannot be expressed by cumulants and
should be measured in order to extract any potential information about the
QCD critical point. We demonstrate that for a wide range of the true
cumulant ratios $K_{n}/K_{m}$ the measured ratios $c_{n}/c_{m}$ quickly
converge if $p<1/2$, which makes the interpretation of the data very
challenging, especially in case of the net-proton cumulants. Our study
further suggests that the measurement of net-charge cumulants may be more
advantageous for the investigation of the QCD phase diagram.
\end{abstract}

\maketitle

\section{Introduction}

The exploration of the phase structure of Quantum Chromodynamics (QCD) has
been a central topic in the study of the strong interactions for many years.
Theoretically, the QCD phase diagram has been investigated with either
effective models or by first principle Lattice QCD calculations.
Experimentally, the phase structure of strongly interacting matter is
studied with the help of relativistic heavy ion collisions. Experiments at
the Relativistic Heavy Ion Collider (RHIC) and recent results from the Large
Hadron Collider (LHC) have indicated that the collisions at the very highest
energies produce a strongly interacting quark-gluon plasma at nearly
vanishing net-baryon density.

Meanwhile Lattice QCD (LQCD) calculations have established that for
vanishing net-baryon density the transition between the hadronic and
partonic ``phase'' is an analytic cross over \cite{Aoki:2006we} with a
pseudo-critical temperature of $T_{c}\simeq 160\,\mathrm{MeV}$ \cite%
{Aoki:2006br,Bazavov:2011nk}. Therefore, true phase transitions, if at all
present, will be located at finite values of the baryon density. Indeed,
model calculations predict a first order coexistence region at high baryon
density (several times nuclear matter density) and moderate temperatures,
which ends in a critical point (for a compilation of various model results
see, e.g., \cite{Stephanov:2007fk}). The precise location of this critical
point, however, is not yet known from theory. Model calculations typically
predict the location of the critical point to be at rather large values of
the baryon number chemical potential, $\mu _{B}$, which currently cannot be 
explored by Lattice QCD methods, due to the fermion sign problem.

Experimentally, the region of the phase diagram at finite density and
moderate temperature is explored by heavy ion collisions at various
energies. Freeze out points extracted from experiments range from values of
nearly vanishing baryon chemical potential for LHC and top RHIC energies to
values of $\mu_{B}\simeq500\,\mathrm{MeV}$ at $\sqrt{s}\simeq5\,\mathrm{GeV}$ %
\cite{BraunMunzinger:2003zd}. Since the location of the critical end point
(CEP) is not really known the entire accessible region of the phase diagram
needs to be explored by scanning the full range of available beam energies.
Both the CEP as well as the first-order phase transition are associated with
characteristic fluctuations -- long range for the second-order transition at
the CEP, and possible spinodal instabilities in case of a first order
transition \cite{Sasaki:2007db,Randrup:2009gp}. Therefore fluctuations of
various quantities such as particle multiplicities or mean transverse
momentum have been proposed as suitable observables \cite{Stephanov:1999zu}.
Measurements of this sort have been \ carried out by the NA49 collaboration %
\cite{Alt:2008ca,:2008vb} at the CERN SPS and are also part of the recently
started RHIC beam energy scan \cite{Aggarwal:2010wy}. The first results by
the NA49 collaboration \cite{Alt:2008ca,:2008vb}, which concentrated on the
variances of various particle ratios and the transverse momentum, showed
only very small deviations from the expected Poisson fluctuations of a
Hadron Resonance Gas (HRG).

Meanwhile it has been realized that higher order cumulants would be more
sensitive to the fluctuations associated with the second order transition,
including the CEP \cite{Stephanov:2008qz,Skokov:2010uh,Stephanov:2011pb}.
The principle reason is that they scale with a higher power of the
correlation length, which will be finite due to the rather short lifetime of
the system created in these collisions. While the cumulants of many
distributions may carry information about the CEP, theoretically the
cumulants of the net-baryon number and of the net-charge are preferable
since they are well defined as derivatives of the QCD partition function
with respect to the appropriate chemical potentials. In addition,
fluctuations of conserved quantities are less affected by final state
interaction in the hadronic phase \cite{Koch:2008ia}. Of course in an actual
experiment, where the baryon number and electric charge of the entire system
are conserved, corrections need to be applied to be able to compare with
theoretical calculations which are typically carried out in the grand
canonical ensemble, where charges are conserved only on the average. This
point has been recently addressed, e.g., in \cite{Bzdak:2012an}.

In addition to corrections due to charge conservation, in a real experiment
one only observes a fraction of the final state particles. In case of the
net-baryon distribution, typically the neutrons are not observed. Furthermore 
there are acceptance cuts as well as efficiency corrections for the
protons, so that less than $50\%$ of baryons are actually observed. As we
shall argue in this paper, this will lead to substantial corrections for the
observed cumulants that need to be taken into account before any
conclusions about critical fluctuations can be drawn. Specifically we will
show that it is \emph{not} sufficient to know the distribution of
net-baryons (or net-charges) $P(N_{B}-N_{\bar{B}})$ in order to predict the
observed cumulants of the net-proton distribution. Instead, information for
the full baryon/anti-baryon distribution, $P(N_{B},N_{\bar{B}})$, is needed,
which is not easily calculated in theory\footnote{%
In principle $P(N_{B},N_{\bar{B}})$ can only be calculated for systems where
a quasi-particle description applies.}. The same is true for the net-charge
distribution, however in this case corrections are smaller since a larger
fraction of charge particles are observed in a typical experiment. On the
positive side, we will show that by measuring various factorial moments of
the observed distribution one is able to extract the cumulants of the
physical net-baryon distribution, $P(N_{B}-N_{\bar{B}})$, which is, after
all, the desired goal.

We note that the question of acceptance corrections to various
  moments of particle distributions have already been discussed in the
  literature. For example corrections to the 
  variance or second order cumulants of the net charge and net baryon
  number distributions have been investigated in 
  Refs. \cite{Bleicher:2000ek,Pruneau:2002yf,Bower:2001fq}, while
  acceptance effects on factorial moments have been addressed in
  \cite{Brooks:1996nu,Kirejczyk:2004sc}. Here we will concentrate on
  the corrections to higher order cumulants which are central to the
  discussion of the QCD critical point. 
  
This paper is organized as follows. In the following section we will derive
the relation between the cumulants of the underlying distribution to that of
the actually measured distribution, and we will show that additional
information other than cumulants is required. In the next section we will
present some examples to illustrate the problems which arise if the
acceptance corrections are large, as it is the case for the measurement of
the net-baryon distribution. Before we conclude we will discuss a few
observations pertinent to our results.

\section{Cumulants}

Suppose we have an underlying probability distribution $P(N_{1},N_{2})$
which captures the full dynamics of the system. Subsequently, we will refer
to $P(N_{1},N_{2})$ as the distribution within the ``required acceptance'', where 
by required acceptance we mean that all
particles that are necessary to capture the relevant physics are measured.
This does not necessarily imply that all particles in the final state need
to be measured.\footnote{%
To which extent an actual experiment does have  
the ``required acceptance'' is not easy to tell and naturally depends on
the specific physics under consideration.} Here $N_{1}$ and $N_{2}$ stand
for either baryons and anti-baryons, i.e., $N_{1}=N_{B}$ and $N_{2}=N_{\bar{B%
}}$ or positively and negatively charged particles, i.e., $N_{1}=N_{+}$ and $%
N_{2}=N_{-}$. Let further $K_{n}$ be the net-baryon or net-charge cumulants
associated with the underlying distribution $P(N_{1},N_{2})$. Of course in a
real experiment additional cuts such as in the transverse momentum or
rapidity will have to be applied. In addition, the efficiency to detect
particles will be smaller that $100\%$ (in case of neutrons it is 
typically zero). To reflect this let us introduce the probability
distribution $p(n_{1},n_{2})$ of the actually observed multiplicities $n_{1}$
and $n_{2}$. Let us further denote the net-baryon or net-charge cumulants of 
$p(n_{1},n_{2})$ by $c_{n}$. Finally, we assume that all acceptance
corrections may be modeled by a binomial probability
distribution\footnote{In reality, more sophisticated models than the
  binomial distribution are likely required to properly address the
  various acceptance corrections (see e.g. \cite{Pruneau:2002yf}). 
  However, using the binomial
  distribution will illustrate the essential issues related with
  acceptance corrections.}
\begin{align}
p(n_{1},n_{2})& =\sum_{N_{1}=n_{1}}^{\infty }\sum_{N_{2}=n_{2}}^{\infty
}P(N_{1},N_{2})\frac{N_{1}!}{n_{1}!(N_{1}-n_{1})!}%
p_{1}^{n_{1}}(1-p_{1})^{N_{1}-n_{1}}  \notag \\
& \times \frac{N_{2}!}{n_{2}!(N_{2}-n_{2})!}%
p_{2}^{n_{2}}(1-p_{2})^{N_{2}-n_{2}}.  \label{pP}
\end{align}

The parameters $p_{1}$ and $p_{2}$ describe all possible acceptance effects
in our system. For example, the detector acceptance is naturally modeled by
the binomial distribution and it can be different for different particles.
Additional cuts in the transverse or/and longitudinal momenta also introduce
acceptance correction. In the context of the cumulants of the net-baryon
distribution, it has been argued in Refs. \cite%
{Kitazawa:2011wh,Kitazawa:2012at} that the fact that neutrons are not
observed in a typical experiment may also be modeled by a binomial
distribution. Thus, Eq. (\ref{pP}) applies as well. Finally, we note that
the different sources of limited acceptance can be represented by an
effective parameter, which is simply the product of all binomial
probabilities. For example, in case of the net-proton distribution: $%
n_{1}=n_{p}$ (number of protons), $n_{2}=n_{\bar{p}}$ (number of
anti-protons), see Eq. (\ref{pP}), and 
\begin{equation}
p_{1}=p_{1}^{\mathrm{B}}\cdot p_{1}^{\mathrm{M}}\cdot p_{1}^{\mathrm{D}}...,
\label{p-mult}
\end{equation}%
where $p_{1}^{\mathrm{B}}$ represents the fact that only protons are
measured instead of baryons ($p_{1}^{\mathrm{B}}\approx 1/2$), $p_{1}^{%
\mathrm{M}}$ characterizes the cut in momentum for measured protons, $p_{1}^{%
\mathrm{D}}$ is a detector efficiency for protons in a given momentum cut,
and so on. Similar relation holds for $p_{2}$.

In order to proceed it is convenient to introduce the factorial moments:%
\begin{align}
F_{ik} & \equiv\left\langle \frac{N_{1}!}{(N_{1}-i)!}\frac{N_{2}!}{(N_{2}-k)!%
}\right\rangle =\sum_{N_{1}=i}^{\infty}\sum_{N_{2}=k}^{\infty}P(N_{1},N_{2})%
\frac{N_{1}!}{(N_{1}-i)!}\frac{N_{2}!}{(N_{2}-k)!},\quad  \label{Fik} \\
f_{ik} & \equiv\left\langle \frac{n_{1}!}{(n_{1}-i)!}\frac{n_{2}!}{(n_{2}-k)!%
}\right\rangle =\sum_{n_{1}=i}^{\infty}\sum_{n_{2}=k}^{\infty}p(n_{1},n_{2})%
\frac{n_{1}!}{(n_{1}-i)!}\frac{n_{2}!}{(n_{2}-k)!}.  \label{fik}
\end{align}
Using Eq. (\ref{pP}) it is straightforward to show (see e.g. \cite{Brooks:1996nu,Kirejczyk:2004sc}):%
\begin{equation}
f_{ik}=p_{1}^{i}\cdot p_{2}^{k}\cdot F_{ik}.  \label{fmg}
\end{equation}
The next step is to express the cumulants $K_{n}$ of the distribution $%
P(N_{1}-N_{2})$ in terms of the factorial moments $F_{ik}$. Using the
relation, Eq. (\ref{fmg}), we can then write $K_{n}$ in terms to the measured
factorial moments $f_{ik}$ for given acceptance parameters $p_{1}$ and $%
p_{2}$. Finally $f_{ik}$ can be expressed by the cumulants $c_{n}$, which
will allow us to relate $K_{n}$ to $c_{m}$. However, as we shall see below,
the factorial moments $f_{ik}$ and $F_{ik}$ in general cannot be solely
expressed in terms of the cumulants of the net distributions, $c_{n}$ and $%
K_{m}$, respectively. Additional terms, not related to cumulants will arise.

It is useful to define the moment generating function:%
\begin{eqnarray}
h(z) & = & \sum\nolimits _{\delta}p(\delta)z^{\delta}  \notag \\
& = & \sum_{N_{1}=0}^{\infty}\sum_{N_{2}=0}^{\infty}P(N_{1},N_{2})\left[%
1-\left(1-z\right)p_{1}\right]^{N_{1}}\left[1-\left(1-z^{-1}\right)p_{2}%
\right]^{N_{2}},  \label{h}
\end{eqnarray}
where $\delta=n_{1}-n_{2}$ {[}see Eq. (\ref{pP}){]} and $p(\delta)$ is the
net multiplicity distribution%
\begin{equation}
p(\delta)=\sum\limits
_{n_{1},n_{2}}p(n_{1},n_{2})\delta_{n_{1}-n_{2}-\delta}.
\end{equation}
Equation (\ref{h}) allows to calculate the cumulant generating function%
\begin{equation}
g(t)=\ln[h(e^{t})]=\sum_{k=1}^{\infty}c_{k}\frac{t^{k}}{k!}.  \label{g}
\end{equation}
For $p_{1}=p_{2}=1$ in Eqs. (\ref{h},\ref{g}) we obtain the moment- and the
cumulant generating function, $G(t)=g(t)|_{p_{1}=p_{2}=1}$, of the net
multiplicity distribution $P(N_{1}-N_{2})$. By definition, the cumulants $%
c_{n}$ and $K_{n}$ read: 
\begin{equation}
c_{n}=\frac{d^{n}g(t)}{dt^{n}}|_{t=0},\quad K_{n}=\frac{d^{n}G(t)}{dt^{n}}%
|_{t=0}\text{.}
\end{equation}

The above equations allow us to calculate the cumulants $c_{n}$ and $K_{n}$.
To obtain the relation between them is straightforward but tedious. For
instance both $c_{2}$ and $K_{2}$ contain $F_{11}$ {[}see Eq. (\ref{Fik}){]}%
, so we express $F_{11}$ by $K_{2}$ and substitute into the expression for $%
c_{2}$. In this way we can relate cumulants $c_{n}$ by cumulants $K_{m}$ and
the factorial moments $F_{ik}$.

First, let us relate the measured cumulants $c_{n}$ with the 
cumulants $K_{m}$ of the underlying
distribution $P\left( N_{1}-N_{2}\right) $. Here we present the results up
to $c_{4}$ and we assume $p=p_{1}=p_{2}$. In the Appendix we give the
general relations up to $c_{6}$ \cite{Friman:2011pf} (see also \cite{Schaefer:2011ex,Gavai:2003mf}) 
with an arbitrary $p_{1}$ and $p_{2}$. We obtain 
\begin{align}
c_{1}& =pK_{1},  \label{c1} \\
c_{2}& =p\left( 1-p\right) N+p^{2}K_{2},  \label{c2} \\
c_{3}& =p(1-p^{2})K_{1}+3p^{2}(1-p)\left( F_{20}-F_{02}-NK_{1}\right)
+p^{3}K_{3},  \label{c3}
\end{align}%
and more complicated:%
\begin{eqnarray}
c_{4} &=&Np(1-p)-3N^{2}p^{2}(1-p)^{2}+6p^{2}(1-p)(F_{02}+F_{20})-\newline
12K_{1}p^{3}(1-p)(F_{20}-F_{02})  \notag \\
&&+6Np^{3}(1-p)(K_{1}^{2}-K_{2})+p^{2}(1-p^{2})(K_{2}-3K_{1}^{2})  \notag \\
&&+6p^{3}(1-p)(F_{03}-F_{12}+F_{02}+F_{20}-F_{21}+F_{30})+p^{4}K_{4}.
\label{c4}
\end{eqnarray}%
To simplify the notation we have introduced%
\begin{equation}
N\equiv \left\langle N_{1}\right\rangle +\left\langle N_{2}\right\rangle
=F_{10}+F_{01}.  \label{N}
\end{equation}%
As can be seen from the above equations it is impossible to relate the
cumulants $%
c_{n}$ solely with the cumulants $K_{m}$ and the parameter $p$. In Eq. (\ref{c2})
the value of $N$ is present. In $c_{3}$ the information about $%
F_{20}=\left\langle N_{1}(N_{1}-1)\right\rangle $ and $F_{02}$ is needed. In
Eq. (\ref{c4}) $F_{30}$, $F_{03}$ and mixed factorial moments $F_{12}$ and $%
F_{21}$ appear. As already discussed in the introduction, these additional
terms would have to be calculated in a given theory, in order to make a
reliable predictions for the measured cumulants $c_{n}$. In general this is
difficult and likely can only be done if a quasi-particle description for
the dynamics applies.

It is interesting to consider two limits of $p$: for $p=1$ we obviously
obtain $c_{n}=K_{n}$, and for very small $p$ the following relations hold%
\begin{equation}
c_{2}\approx\left\langle n_{1}\right\rangle +\left\langle n_{2}\right\rangle
,\quad c_{3}\approx c_{1},\quad c_{4}\approx c_{2},
\end{equation}
where $\left\langle n_{i}\right\rangle =p\left\langle N_{i}\right\rangle $.
In the limit $p\rightarrow0$ the ratios of cumulants read: $c_{3}/c_{1}=1$, $%
c_{4}/c_{2}=1$. It means that even if the cumulant ratio $K_{n}/K_{m}$
contains information about the QCD phase diagram (or any other interesting
physics) this information is lost if the acceptance is too small. In other
words in the limit $p\rightarrow0$ the Skellam distribution is obtained,
i.e., we are in the limit of Poisson statistics.

To summarize, Eqs. (\ref{c1}-\ref{c4}) allow to obtain the values of
cumulants $c_{n}$ at a given acceptance parameter $p$, knowing the cumulants 
$K_{n}$ and certain factorial moments $F_{ik}$ of the underlying
distribution $P\left( N_{1,}N_{2}\right) $. This distribution according to
our assumptions is obtained for the ``required acceptance'', defined by $%
p=1$.

In practice, the more interesting question is to which extent one can
express the cumulants $K_{n}$, which contain the relevant physics, in terms
of the measurable cumulants, $c_{m}$, and factorial moments, $f_{ik}$.
Actually this can be easily done using Eqs. (\ref{c1}-\ref{c4}) and the
relation (\ref{fmg}):%
\begin{align}
pK_{1}& =c_{1},  \label{K1} \\
p^{2}K_{2}& =c_{2}-n(1-p),  \label{K2} \\
p^{3}K_{3}& =c_{3}-c_{1}(1-p^{2})-3(1-p)(f_{20}-f_{02}-nc_{1}),  \label{K3}
\end{align}%
and%
\begin{eqnarray}
p^{4}K_{4}
&=&c_{4}-np^{2}(1-p)-3n^{2}(1-p)^{2}-6p(1-p)(f_{20}+f_{02})+12c_{1}(1-p)(f_{20}-f_{02})
\notag \\
&&-(1-p^{2})(c_{2}-3c_{1}^{2})-6n(1-p)(c_{1}^{2}-c_{2})  \notag \\
&&-6(1-p)(f_{03}-f_{12}+f_{02}+f_{20}-f_{21}+f_{30}).  \label{K4}
\end{eqnarray}%
where we introduced the notation%
\begin{equation}
n\equiv \left\langle n_{1}\right\rangle +\left\langle n_{2}\right\rangle
=f_{10}+f_{01}.
\end{equation}

These equations can be directly used to extract the values of the true
cumulants $K_{n}$ that characterize the system with $p=1$, which we
assumed to capture the relevant physics. We repeat: The measurement of the
cumulants $c_{m}$ is \emph{not} sufficient to extract the desired cumulants $%
K_{n}$. The additional measurement of various factorial moments $f_{ik}$ is
required as well.

For example, in the context of the net-baryon fluctuations, $K_{n}$ denote
the net-baryon cumulants. To determine $K_{n}$ all we need as input is the
value of the acceptance parameter $p$, the measured cumulants $c_{n}$ of the
net-proton distribution, and the factorial moments $f_{ik}$ measured for
protons and anti-protons.

\section{Examples}

In the following we present a few examples where we show the relation
between the observed cumulants $c_{n}$ and the cumulants of the underlying
distribution, $K_{n}$. These are merely examples to illustrate the situation
and certain assumptions about the necessary factorial moments will have to
be made.

\subsection{$c_{3}/c_{1}$}

As an example we show the dependence of the ratio $c_{3}/c_{1}$ as a
function of the acceptance parameter $p$ for different values of $%
K_{3}/K_{1} $. As seen from Eq. (\ref{c3}), $c_{3}$ depends only on
cumulants $K_{1}$ and $K_{3}$ and factorial moments $F_{20}$ and $F_{02}$.
To proceed we denote%
\begin{equation}
\left\langle N_{i}^{2}\right\rangle -\left\langle N_{i}\right\rangle
^{2}=\left\langle N_{i}\right\rangle (1+\alpha ),  \label{alpha}
\end{equation}%
where $i=1,2$ and $\alpha $ is a free parameter that allows to change the
widths of the distributions $P(N_{1})$ and $P\left( N_{2}\right) $ \footnote{%
For simplicity we assume that $\alpha $ is the same for $N_{1}$ and $N_{2}$
multiplicity distributions.}. Substituting Eq. (\ref{alpha}) to Eq. (\ref{c3}%
) and using Eq. (\ref{c1}), we obtain:%
\begin{equation}
\frac{c_{3}}{c_{1}}=1-p^{2}+3\alpha p(1-p)+p^{2}\frac{K_{3}}{K_{1}}.
\end{equation}

\begin{figure}[h]
\begin{centering}
\includegraphics[scale=0.4]{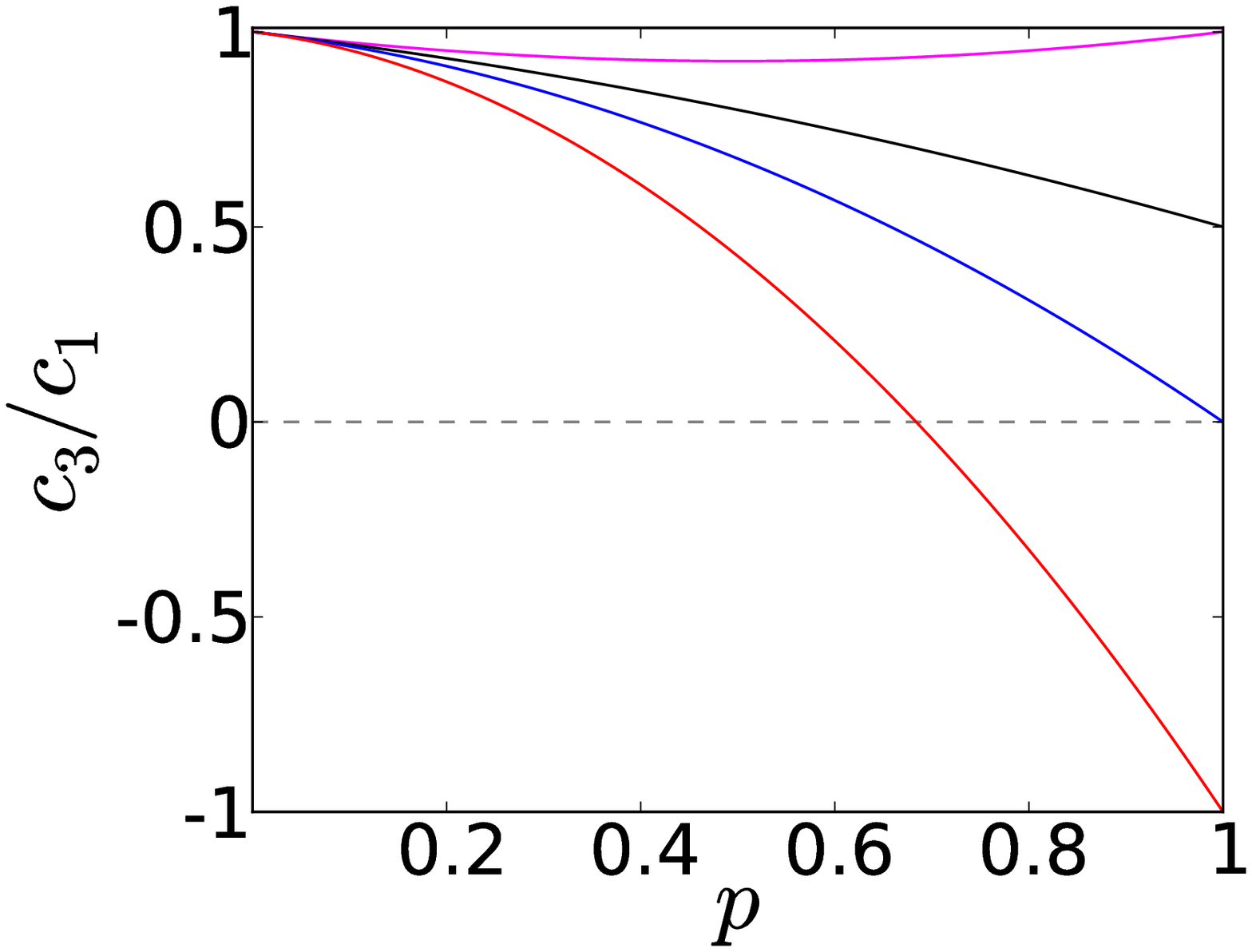} \hspace{1.0cm} 
\includegraphics[scale=0.4]{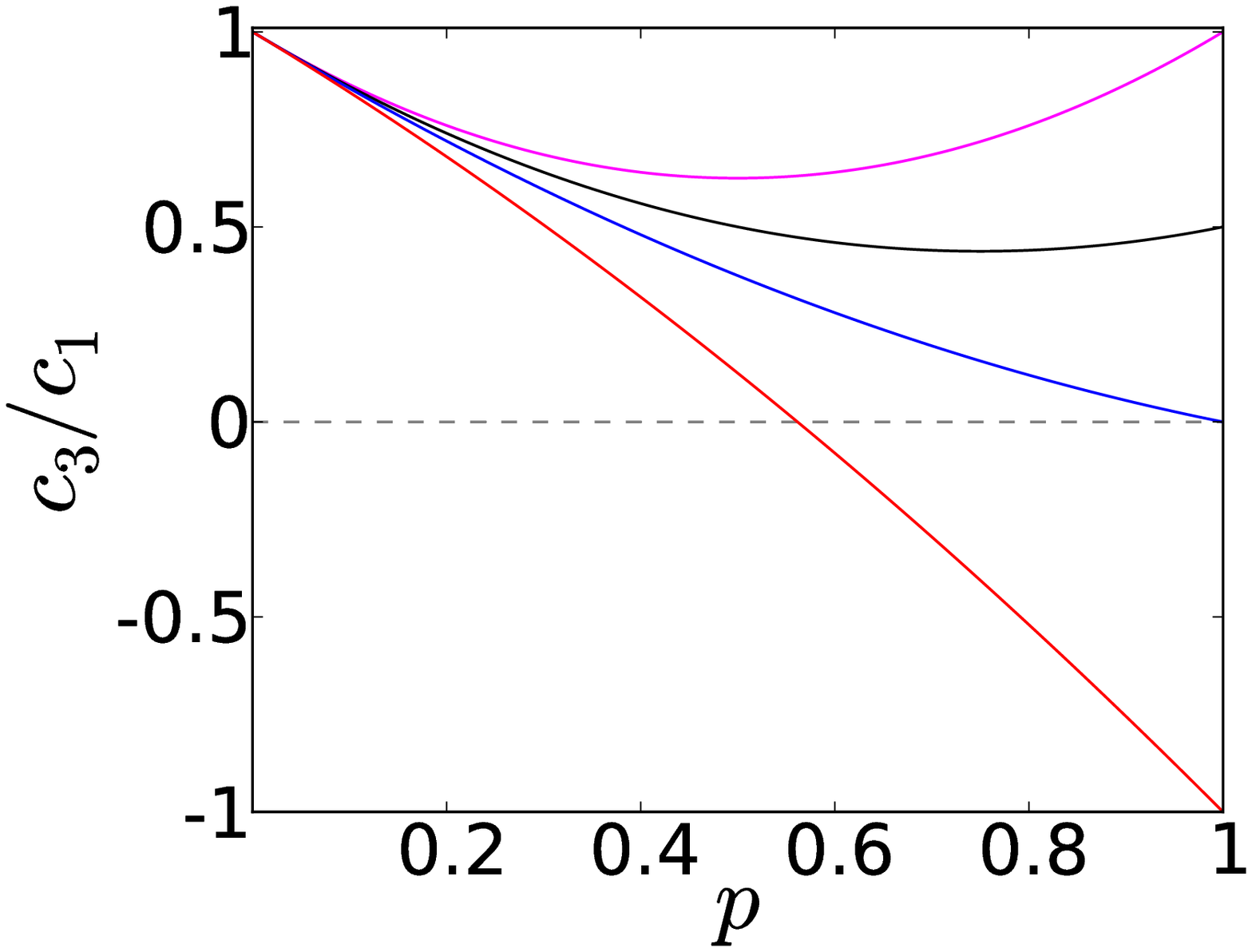} 
\par\end{centering}
\caption{The measured cumulant ratio $c_{3}/c_{1}$ as a function of the
acceptance parameter $p$ for four values of $K_{3}/K_{1}=-1$, $0$, $0.5$ and 
$1$. $K_{3}/K_{1}$ equals $c_{3}/c_{1}$ at $p=1$. In the left plot $\protect%
\alpha =-0.1$ and in the right plot $\protect\alpha =-0.5$, see Eq. (\ref%
{alpha}). For the measurement of the net-proton cumulants at STAR the realistic
value of $p$ is smaller than $1/2$ and arguably close to $1/5$.}
\label{fig_1}
\end{figure}

\begin{figure}[h]
\begin{centering}
\includegraphics[scale=0.4]{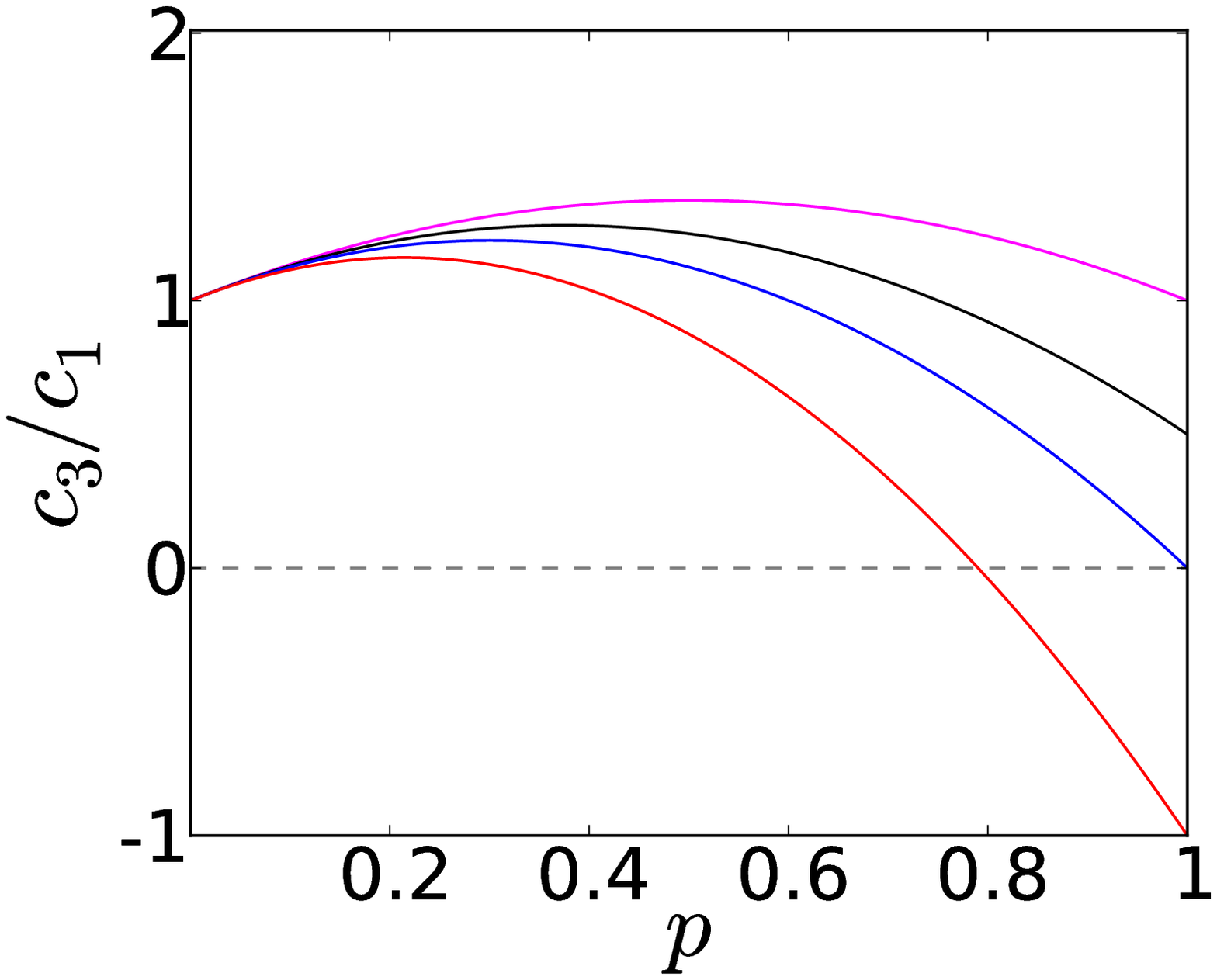} \hspace{1.0cm} 
\includegraphics[scale=0.4]{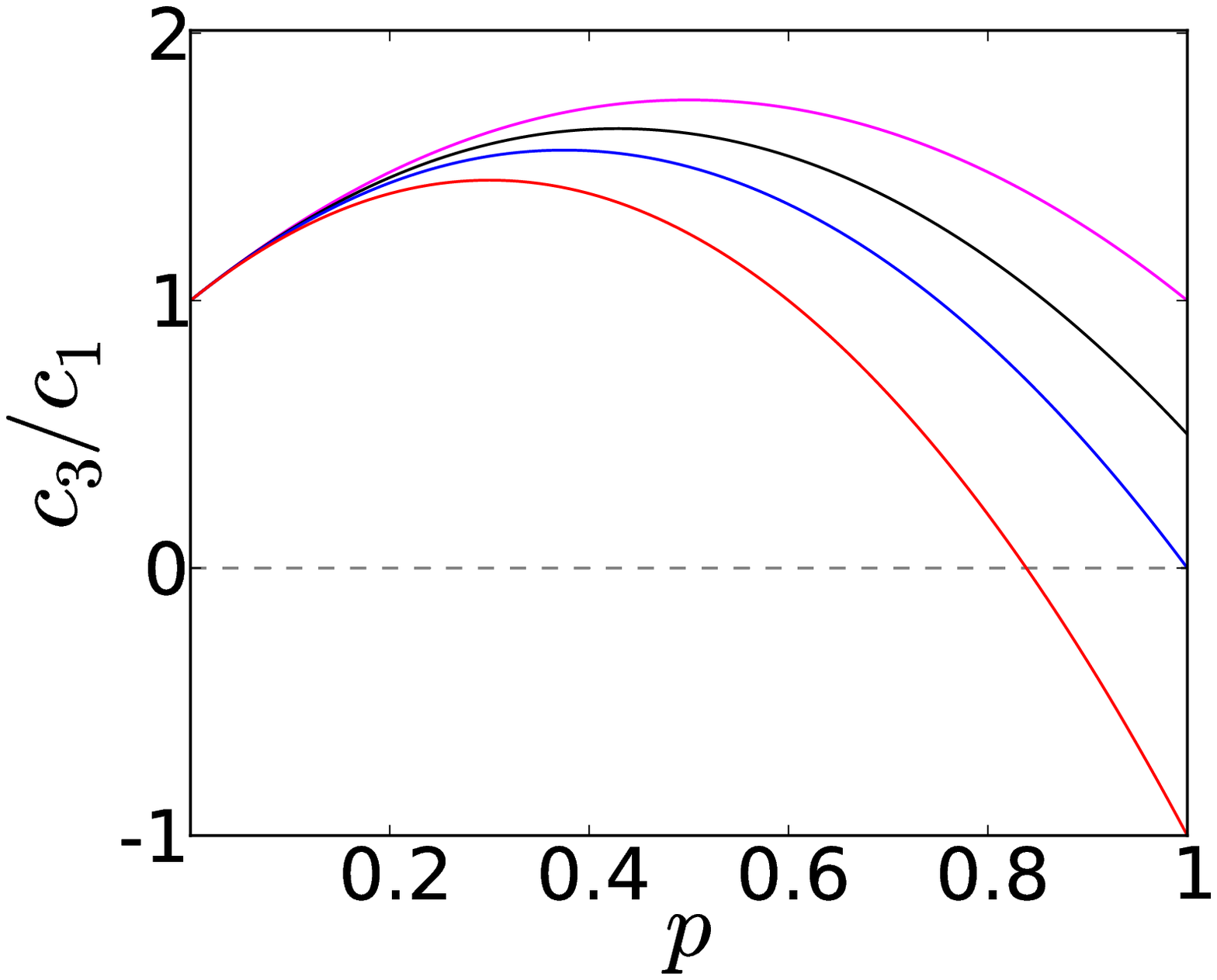} 
\par\end{centering}
\caption{The same as in Fig. \ref{fig_1}, except in the left plot $\protect%
\alpha=0.5$ and in the right plot $\protect\alpha=1$, see Eq. (\ref{alpha}).}
\label{fig_2}
\end{figure}

In the context of the net-baryon fluctuations it is expected that the width
of the baryon (anti-baryon) distribution is comparable to the Poisson
distribution \cite{Aggarwal:2010wy}, that is $|\alpha |<<1$. However, when
 net-charge fluctuations are studied the parameter $\alpha $ is expected
to be of the order of one \footnote{%
The charge multiplicity distribution in heavy-ion and proton-proton
collisions can be described by the negative binomial distribution, see,
e.g., \cite{Adare:2008ns}.}. In Fig. \ref{fig_1} and Fig. \ref{fig_2} we
present the dependence of $c_{3}/c_{1}$ as a function of $p$ for four values
of $K_{3}/K_{1}=-1,$ $0$, $0.5$ and $1$, and we note that, as previously
discussed, $K_{3}/K_{1}$ equals to the value of $c_{3}/c_{1}$ at $p=1$. In
Fig. \ref{fig_1} we assume that the multiplicity distribution is narrower
than the Poisson distribution: $\alpha =-0.1$ in the left plot, and $\alpha
=-0.5$ in the right plot. 

In Fig.~\ref{fig_2} we choose $\alpha$ to be positive, i.e., the
multiplicity distribution is broader than the Poisson distribution: $%
\alpha=0.5$ in the left plot, and $\alpha=1$ in the right plot. 

In case of net-baryon fluctuations the fact that we only measure protons
(anti-protons) instead of all baryons (anti-baryons) already introduces $%
p\lesssim1/2$. In addition finite detection efficiencies and cuts in the
transverse momentum reduce the value of $p$ even more. We estimate that for
the STAR measurement \cite{Aggarwal:2010wy} the parameter $p\approx1/5$ may
not be far from reality (assuming that the cut in rapidity is good enough to
capture the relevant physics, and its contribution to $p$ can be neglected).
As seen in Figs. \ref{fig_1} and \ref{fig_2} already the value of $p=1/2$
shifts the observed ratio $c_{3}/c_{1}$ into positive values even if $%
K_{3}/K_{1}$ is negative. Moreover, for different values of $K_{3}/K_{1}$
the values of $c_{3}/c_{1}$ quickly converge which makes the interpretation
of the experimental results challenging. Taking a more realistic value of $%
p=1/5$ we see that it is practically impossible to distinguish between
different values of $K_{3}/K_{1}$, unless $K_{3}/K_{1}<<-1$.

\subsection{$c_{4}/c_{2}$}

It is interesting to repeat the same exercise for $c_{4}/c_{2}$. However, in
this case we need additional information about $F_{30}$, $F_{03}$ and the
mixed factorial moments $F_{21}$ and $F_{12}$. We investigated different
values of parameters and we obtain similar plots as in Figs. \ref{fig_1} and %
\ref{fig_2}. To present a few examples, we will assume that $F_{ik}=F_{ki}$ 
\footnote{%
We note that this choice corresponds to a system at zero baryon chemical
potential in the context of the net-baryon cumulants. However, the formalism
presented in this paper is not restricted to this choice. Also, the
corrections shown here are expected to be similar in case of finite
chemical potential.}, and introduce the following parametrization%
\begin{eqnarray}
F_{20} &=&\left\langle N_{1}\right\rangle ^{2}+\alpha \left\langle
N_{1}\right\rangle ,  \label{al-F20} \\
F_{30} &=&\left\langle N_{1}\right\rangle ^{3}+3\alpha \left\langle
N_{1}\right\rangle ^{2}+2\alpha ^{2}\left\langle N_{1}\right\rangle ,
\label{al-F30} \\
F_{21} &=&F_{20}\left\langle N_{1}\right\rangle +\beta \left\langle
N_{1}\right\rangle ^{2}.  \label{b-F21}
\end{eqnarray}%
Equations (\ref{al-F20},\ref{al-F30}) result from the assumption that $%
P(N_{1})$ is given by the negative binomial distribution. The parameter $%
\alpha $ is identical as in Eq. (\ref{alpha}), and controls the width of the
multiplicity distribution. $\beta $ reflects the correlation between $N_{1}$
and $N_{2}$. To further reduce the number of parameters we assume that 
\footnote{%
We checked that modifications of this assumption do not introduce any new
qualitative features.}: 
\begin{equation}
K_{2}=2\left\langle N_{1}^{2}\right\rangle -2\left\langle
N_{1}N_{2}\right\rangle \approx 2\left\langle N_{1}^{2}\right\rangle
-2\left\langle N_{1}\right\rangle ^{2}.
\end{equation}%
Finally, the value of $K_{4}$ is determined from the ratio $K_{4}/K_{2}$
that is an input in our calculation.

\begin{figure}[h]
\begin{centering}
\includegraphics[scale=0.4]{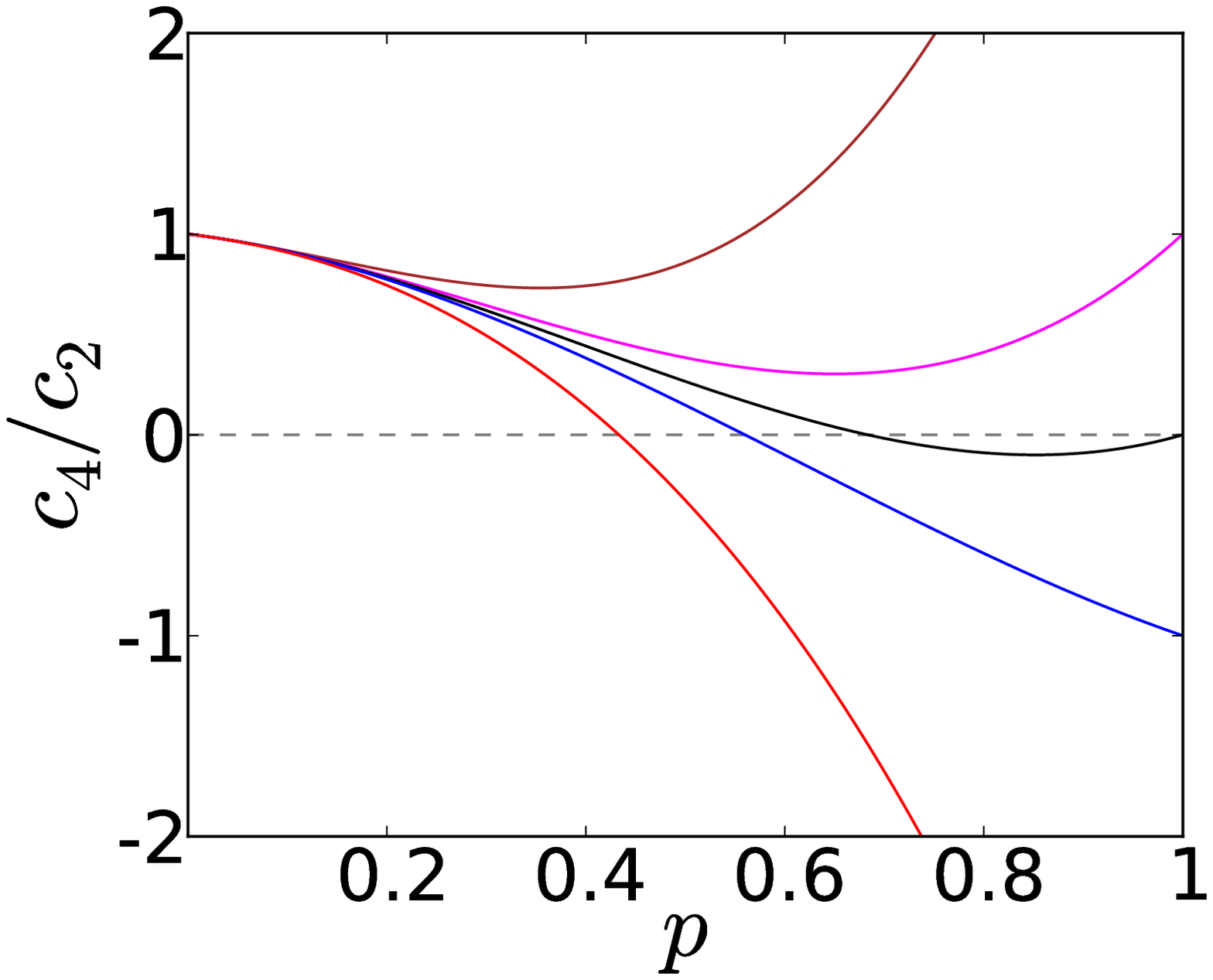} \hspace{1.0cm} 
\includegraphics[scale=0.4]{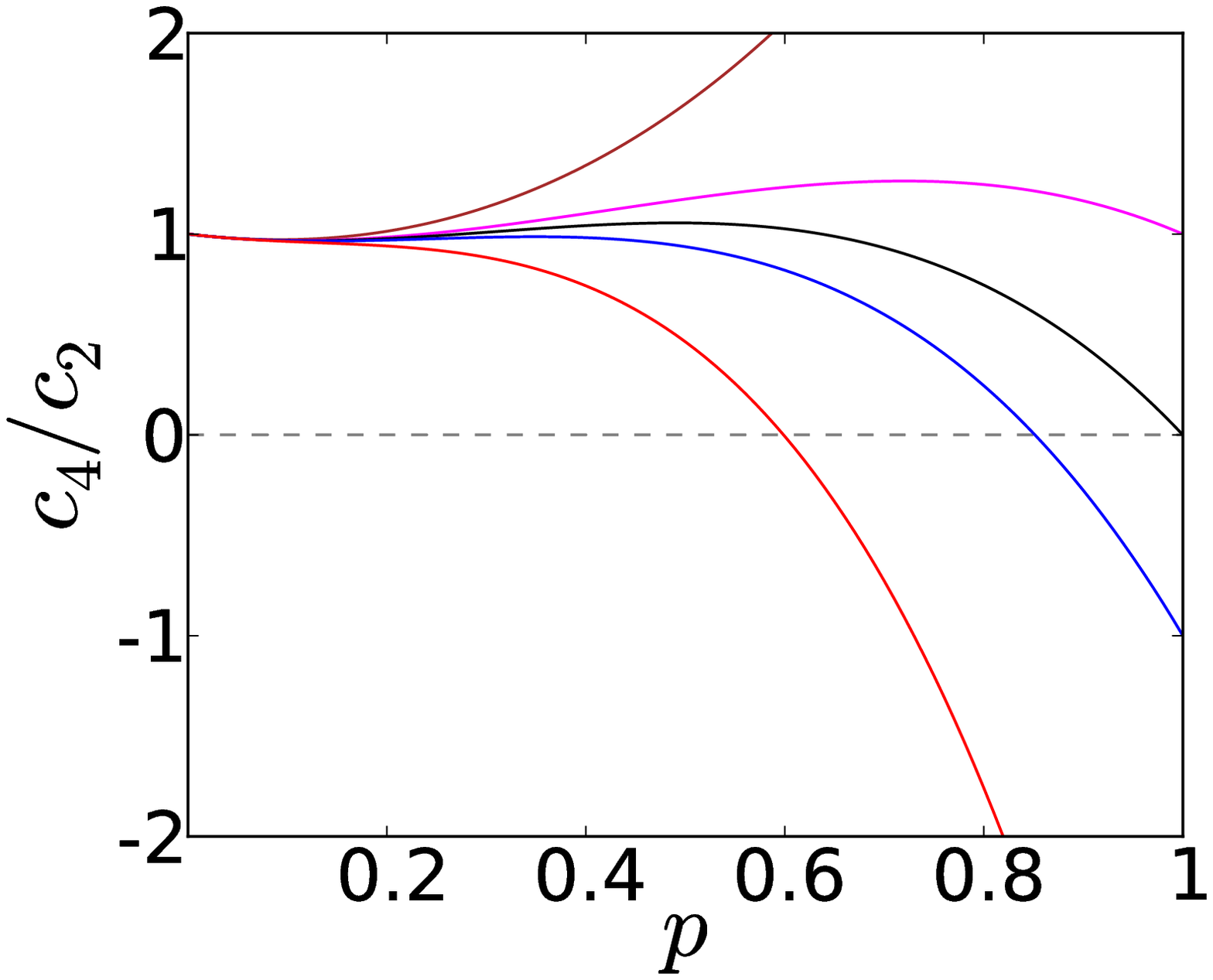} 
\par\end{centering}
\caption{The measured cumulant ratio $c_{4}/c_{2}$ as a function of the
acceptance parameter $p$ for five values of $K_{4}/K_{2}=-5$, $-1$,
$0,$ $1,$ and $5$. $K_{4}/K_{2}$ equals $c_{4}/c_{2}$ at $p=1$. In both plots $\protect%
\alpha =-0.1$. In the left plot $\protect\beta =0.01$, and in the right one $%
\protect\beta =-0.01$, see Eqs. (\ref{al-F20}-\ref{b-F21}). For the measurement
of the net-proton cumulants at STAR the realistic value of $p$ is smaller
than $1/2$ and arguably close to $1/5$.}
\label{fig_3}
\end{figure}

\begin{figure}[h]
\begin{centering}
\includegraphics[scale=0.4]{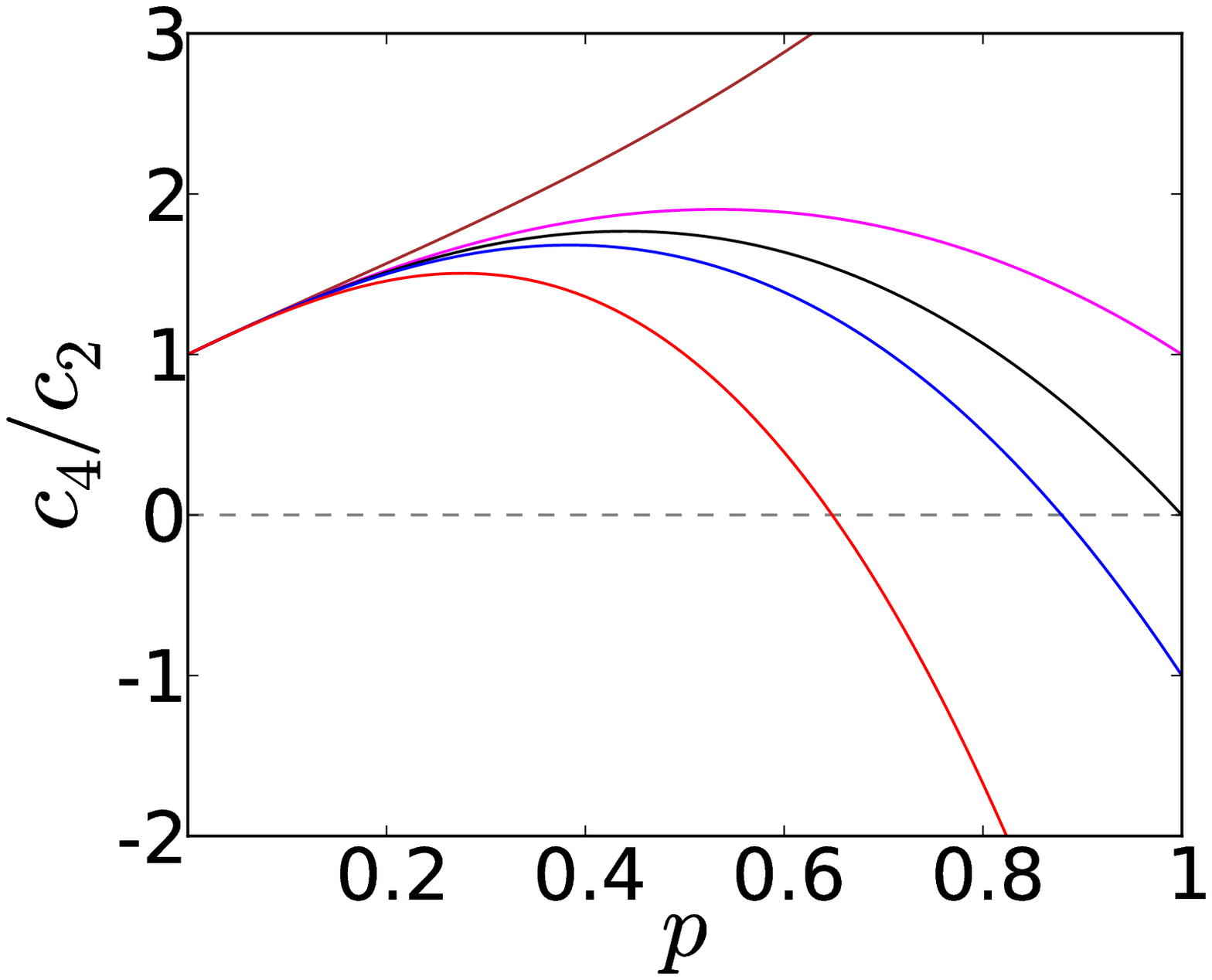} \hspace{1.0cm} 
\includegraphics[scale=0.4]{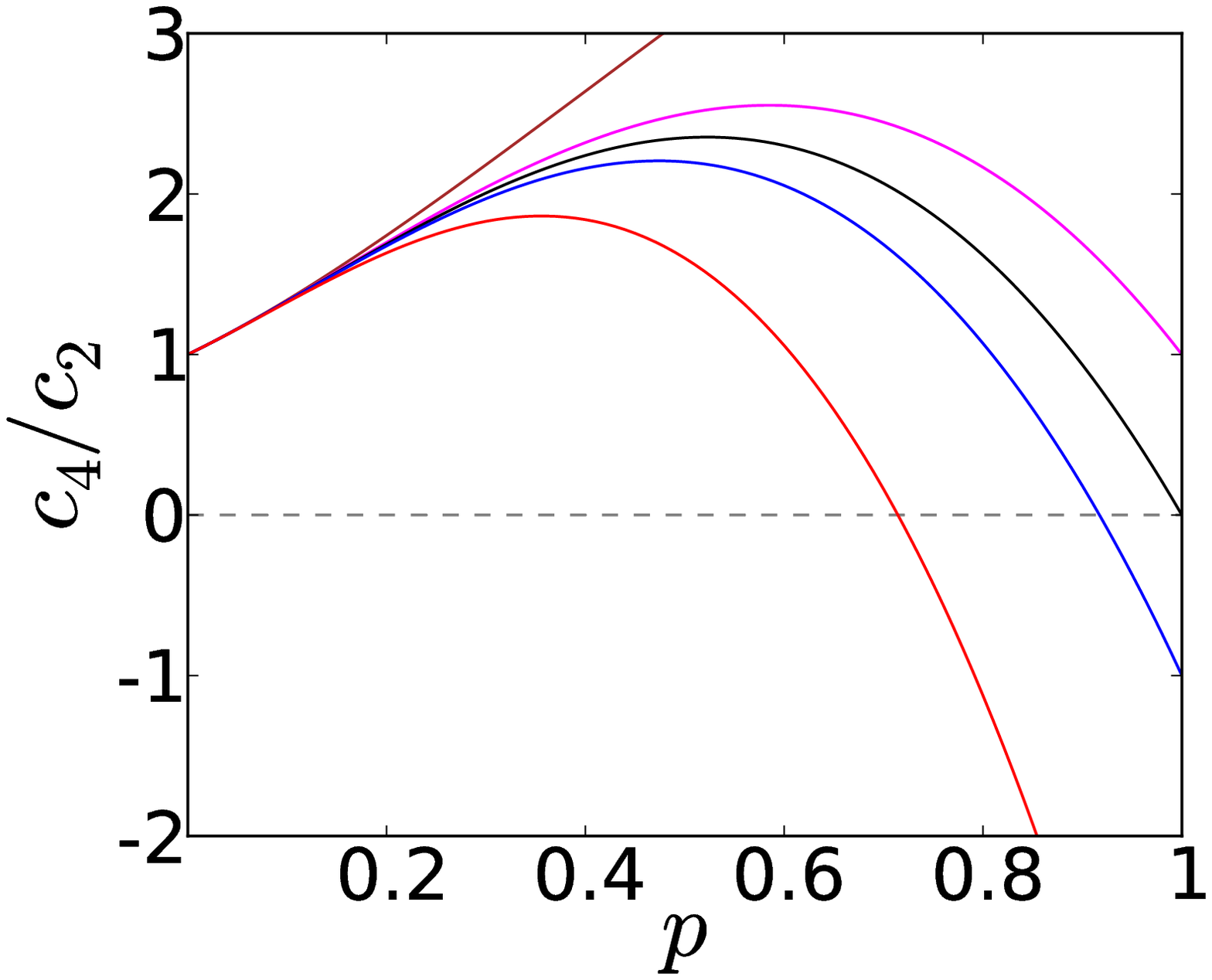} 
\par\end{centering}
\caption{The same as in Fig. \ref{fig_3}, except in the left plot $\protect%
\alpha=0.5$, $\protect\beta=0.01$, and in the right plot $\protect\alpha=0.5$%
, $\protect\beta=-0.01$. }
\label{fig_4}
\end{figure}

In Fig.~\ref{fig_3} and Fig.~\ref{fig_4} we present the dependence of $%
c_{4}/c_{2}$ as a function of the acceptance parameter $p$ for five values
of $K_{4}/K_{2}=-5,$ $-1,$ $0,$ $1,$ and $5$. In Fig. \ref{fig_1} we assume $N=100$, 
see Eq. (\ref{N}), and $\alpha =-0.1$. In the left plot $\beta =0.01$, and
in the right one $\beta =-0.01$. 

In Fig. \ref{fig_4} we assume $N=100$ and $\alpha=0.5$. In this case the
multiplicity distribution is broader than Poisson, which is expected if the
net-charge cumulants are investigated. In the left plot $\beta=0.01$, and in
the right one $\beta=-0.01$. 

As seen from Figs. \ref{fig_3} and \ref{fig_4} the convergence of $%
c_{4}/c_{2}$ for various values of $K_{4}/K_{2}$ is even more rapid than in
case of $c_{3}/c_{1}$. Taking $p=1/5$, as in the STAR measurement of
net-proton cumulants, we see that even if there is any information about the
QCD phase diagram in the net-baryon cumulants, this information is strongly
diluted if only protons and anti-protons are measured.\footnote{Our
finding naturally explains the large difference between net-baryon and
net-proton cumulants observed in the UrQMD model \cite{Schuster:2009jv}.}
At $p=1/5$ all lines are practically indistinguishable, unless 
$|K_{4}/K_{2}|>>5$ (it should be of the order of $50$).

\section{Comments}

Let us discuss a few observations pertinent to the results from the previous
sections:

(i) We demonstrated that a limited acceptance \footnote{%
In particular the difficulty in measuring neutrons in the context of
net-baryon fluctuations.} makes a direct interpretation of measurements of
higher order cumulants very difficult. Instead, we propose that the
measurement of additional factorial moments, as shown in Eqs. (\ref{K1}-\ref%
{K4}), or the more general ones in the Appendix, is the way to proceed. This
allows a rather unambiguous determination of the true cumulants $K_{n}$.

(ii) The main problem with the measurement of the net-proton fluctuations is
that the maximal value of $p$ is approximately $1/2$.\footnote{Of
  course if one compares with theoretical predictions which directly
  calculate the proton cumulants, such as \cite{Athanasiou:2010kw},
  the effective binomial probability $p$ is closer to one, similar to that
  for charged particles.  However, it is not clear
  how to determine the proton cumulants in a model independent way.} Moreover, additional
cuts in the transverse momentum are usually needed. Also detection
efficiencies are never $100\%$. As a result, $p<1/2$ which makes the
interpretation of the data very challenging. Even if one measured the
additional factorial moments as proposed in this paper, their contribution
becomes large for small values of $p$ {[}see Eqs. (\ref{K1}-\ref{K4}){]} and
thus need to be determined very precisely. This clearly favors the analysis
of net-charge fluctuations, where there is no problem equivalent to the
neutrons. Thus the parameter $p$ will likely be larger than $1/2$ and the
extrapolation to the full acceptance by measuring the necessary factorial
moments will be much more reliable. The cumulants of the net-charge
fluctuations should also be sensitive to the CEP, however with a potentially
smaller overall magnitude \cite{Stephanov:2008qz}.

(iii) For the purpose of this paper, we have assumed that the ``required
acceptance'', i.e., $p=1$, captures the relevant physics. To which extent
this is the case for an actual experiment, such as STAR, is not clear and
difficult to estimate. For example, it is not clear to us what the size of
the rapidity window should be, in order to capture the relevant physics. One
estimator would be the width of a thermal fireball, which is of the order of
one unit of rapidity. However, it may very well be larger, and, therefore,
the \textit{true} value of $p$ may very well be smaller than $1/5$ in the
STAR measurement.

(iv) Given the expression for the true cumulants $K_{n}$ in terms of the
measured cumulants $c_{m}$ and factorial moments, Eqs. (\ref{K1}-\ref{K4}),
it may be conceivable that one could explore the necessary rapidity range
experimentally. As one increases the rapidity window the resulting $K_{n}$
should approach an asymptotic value once the window captures the relevant
physics. On the other hand, doing the same exercise with the measured
cumulants $c_{n}$ only may lead to false conclusions, especially in
case of the 
net-proton distribution. Since the binomial parameter $p$ is already very
small reducing the rapidity window will only lead to small variations (see
Figs. \ref{fig_1}-\ref{fig_4}) suggesting that the physics does not change.

(v) In our examples for $c_{4}/c_{2}$ we find that the choice for the correlation
term $\sim \beta $ in Eq. (\ref{b-F21}) leads to significant corrections for
the measured cumulant ratio. Therefore, these correlation terms need to be
measured precisely and we expect them to play an even stronger role in
higher order cumulant ratios, such as $c_{6}/c_{2}$.

(vi) We note the formalism presented here may be applied to higher order
cumulants, such as $K_{6}$, in a straightforward but tedious way
(see Appendix). We believe that the problem to resolve the true cumulants gets
even more difficult, the higher the order of the cumulant under
consideration is.

(vii) Instead of measuring the cumulants one may consider other
moments which are independent of the (binomial)
acceptance corrections. For example  the scaled factorial moments 
\cite{Pruneau:2002yf,Brooks:1996nu,Kirejczyk:2004sc},
$f_{ik}^{(s)}=\frac{f_{ik}}{\ave{n_1}^i\ave{n_2}^k}=\frac{F_{ik}}{\ave{N_1}^i\ave{N_2}^k}$,
see Eqs. (\ref{Fik}, \ref{fik}), would be such an alternative. Or one
could construct combinations of moments similar to the one suggested
in  \cite{Bower:2001fq} for the second moments.   
While these alternative  moments should in principle carry
similar information as the cumulants, they are not fully determined by
the net-baryon distribution, $P(N_B-N_{\bar{B}})$. Therefore, a
theoretical determination of these moments, e.g. from Lattice QCD,
will necessarily require some model-dependent assumptions.

\section{Conclusions}

In this paper we have studied the effect of finite acceptance on higher
order cumulants of net-baryon and net-charge distributions. To this end we
have folded the true probability distribution with binomial
distribution in order to simulate the finite acceptance corrections. Our
main finding is that the task of extracting the cumulants $K_{n}$ of the
true distribution requires the  measurement of not only the
cumulants $c_{n}$ of the measurable distribution but in addition of
various factorial moments, which \emph{cannot} be expressed in terms
of cumulants. 
We also demonstrated
that for various values of the true cumulants ratios $K_{n}/K_{m}$ the
measured ratios $c_{n}/c_{m}$ quickly converge with decreasing acceptance
parameter $p$. This makes the physical interpretation of the net-proton
cumulants very difficult. We further argued that it may be advantageous to
investigate the cumulants and factorial moments of the net-charge
distribution, since in this case acceptance corrections are considerably
smaller than for the net-baryon distribution. This may allow for a more
reliable extraction of the true cumulants via the methods presented in this
paper.

\bigskip

\section{Acknowledgments}

A.B. was supported by Contract No. DE-AC02-98CH10886 with the U. S.
Department of Energy. V.K. was supported by the Office of Nuclear Physics in
the US Department of Energy's Office of Science under Contract No.
DE-AC02-05CH11231. A.B. also acknowledges the grant N N202 125437 of the
Polish Ministry of Science and Higher Education (2009-2012).

\appendix       

\section{General relations}

Here we present the general relations between cumulants $K_{n}$
characterized by the ``required acceptance'', and the measurable
factorial moments at a given acceptance parameters $p_{1}$ and $p_{2}$. As
seen in Eqs. (\ref{K2}-\ref{K4}) it is not possible to express cumulants $%
K_{n}$ solely by cumulants $c_{m}$ but also factorial moments $f_{ik}$
appear. Therefore, here we express $K_{n}$ solely by the factorial moments $%
f_{ik}$. This significantly simplifies our notation but makes no difference
for an experimental application. We obtain 
\begin{align}
K_{1}& =\left\langle N_{1}\right\rangle -\left\langle N_{2}\right\rangle , \\
K_{2}& =N-K_{1}^{2}+F_{02}-2F_{11}+F_{20}, \\
K_{3}& =K_{1}+2K_{1}^{3}-F_{03}-3F_{02}+3F_{12}+3F_{20}-3F_{21}+F_{30} 
\notag \\
& -3K_{1}(N+F_{02}-2F_{11}+F_{20}), \\
K_{4}& =N-6K_{1}^{4}+F_{04}+6F_{03}+7F_{02}-2F_{11}-6F_{12}-4F_{13}  \notag
\\
& +7F_{20}-6F_{21}+\newline
6F_{22}+6F_{30}-4F_{31}+F_{40}  \notag \\
& +\newline
12K_{1}^{2}(N+F_{02}-2F_{11}+F_{20})-3(N+F_{02}-2F_{11}+F_{20})^{2}  \notag
\\
& -4K_{1}(K_{1}-F_{03}-3F_{02}+3F_{12}+3F_{20}-3F_{21}+F_{30}),
\end{align}%
and $K_{5}$ and $K_{6}$ are more complicated%
\begin{align}
K_{5}&
=K_{1}+24K_{1}^{5}-F_{05}-10F_{04}-25F_{03}-15F_{02}+15F_{12}+20F_{13}+5F_{14}
\notag \\
& +15F_{20}-\newline
15F_{21}-10F_{23}+25F_{30}-\newline
20F_{31}+10F_{32}+10F_{40}-5F_{41}+F_{50}  \notag \\
& -\newline
60K_{1}^{3}(N+F_{02}-2F_{11}+F_{20})+30K_{1}(N+F_{02}-2F_{11}+F_{20})^{2} 
\notag \\
& +20K_{1}^{2}(K_{1}-F_{03}-3F_{02}+3F_{12}+3F_{20}-3F_{21}+F_{30})  \notag
\\
& -\newline
10(N+F_{02}-2F_{11}+F_{20})(K_{1}-F_{03}-3F_{02}+3F_{12}+3F_{20}-3F_{21}+F_{30})
\notag \\
& -5K_{1}\newline
(N+F_{04}+6F_{03}+7F_{02}-2F_{11}-6F_{12}-4F_{13}+7F_{20}-6F_{21}+6F_{22} 
\notag \\
& +6F_{30}-4F_{31}+F_{40}),
\end{align}%
and%
\begin{align}
K_{6}& =N-120K_{1}^{6}+F_{06}+15F_{05}+65F_{04}+90F_{03}+31F_{02}-2F_{11}-%
\newline
30F_{12}-80F_{13}  \notag \\
& -45F_{14}-6F_{15}+31F_{20}-30F_{21}+30F_{22}+\newline
30F_{23}+15F_{24}+90F_{30}-80F_{31}  \notag \\
& +30F_{32}-20F_{33}+65F_{40}-\newline
45F_{41}+15F_{42}+15F_{50}-6F_{51}+F_{60}  \notag \\
& +360K_{1}^{4}(N+F_{02}-2F_{11}+F_{20})-\newline
270K_{1}^{2}(N+F_{02}-2F_{11}+F_{20})^{2}  \notag \\
&
+30(N+F_{02}-2F_{11}+F_{20})^{3}-120K_{1}^{3}(K_{1}-F_{03}-3F_{02}+3F_{12}+3F_{20}
\notag \\
& -3F_{21}+F_{30})+\newline
120K_{1}(N+F_{02}-2F_{11}+F_{20})(K_{1}-F_{03}-3F_{02}+3F_{12}  \notag \\
& +3F_{20}-3F_{21}+F_{30})-\newline
10(K_{1}-F_{03}-3F_{02}+3F_{12}+3F_{20}-3F_{21}+F_{30})^{2}  \notag \\
& +\newline
30K_{1}^{2}(N+F_{04}+6F_{03}+7F_{02}-2F_{11}-6F_{12}-4F_{13}+7F_{20}-6F_{21}
\notag \\
& +6F_{22}+6F_{30}-4F_{31}+F_{40})-\newline
15(N+F_{02}-2F_{11}+F_{20})\newline
(N+F_{04}+6F_{03}  \notag \\
&
+7F_{02}-2F_{11}-6F_{12}-4F_{13}+7F_{20}-6F_{21}+6F_{22}+6F_{30}-4F_{31}+F_{40})
\notag \\
& -6K_{1}(K_{1}-F_{05}-10F_{04}-25F_{03}-15F_{02}+15F_{12}+\newline
20F_{13}+5F_{14}+15F_{20}  \notag \\
& -15F_{21}-10F_{23}+25F_{30}-20F_{31}+10F_{32}+10F_{40}-5F_{41}+F_{50}).
\end{align}%
In the above equations all cumulants $K_{n}$ can be directly measured due to
the equality:%
\begin{equation}
F_{ik}=\frac{1}{p_{1}^{i}p_{2}^{k}}f_{ik},
\end{equation}%
and 
\begin{align}
N& =\left\langle N_{1}\right\rangle +\left\langle N_{2}\right\rangle =\frac{%
\left\langle n_{1}\right\rangle }{p_{1}}+\frac{\left\langle
n_{2}\right\rangle }{p_{2}}, \\
K_{1}& =\left\langle N_{1}\right\rangle -\left\langle N_{2}\right\rangle =%
\frac{\left\langle n_{1}\right\rangle }{p_{1}}-\frac{\left\langle
n_{2}\right\rangle }{p_{2}}.
\end{align}

\end{document}